\input{aipcheck}

\documentclass[
    ,final            
  ,numberedheadings 
  ,nomathfonts                 
  ]
  {aipproc}

\layoutstyle{6x9}

\newcommand{\be}{\begin{equation}}
\newcommand{\ee}{\end{equation}}
\newcommand{\beq}{\begin{eqnarray}}
\newcommand{\eeq}{\end{eqnarray}}
\newcommand{\beqno}{\begin{eqnarray*}}
\newcommand{\eeqno}{\end{eqnarray*}}
\newcommand{\bitmz}{\begin{itemize}}
\newcommand{\eitmz}{\end{itemize}}

\newcommand{\ltsim}{\raisebox{-0.6ex}{$\,\stackrel
        {\raisebox{-.2ex}{$\textstyle <$}}{\sim}\,$}}
\newcommand{\gtsim}{\raisebox{-0.6ex}{$\,\stackrel
        {\raisebox{-.2ex}{$\textstyle >$}}{\sim}\,$}}

\def\Ledd{L_{\rm Edd}}

\def\Tin{T_{\rm in}}

\def\epiv{E_{\rm p}}

\def\chired{\chi^2_{\rm red}}
\def\ergs{{\rm erg\,s^{-1}}}

\def\araa{ARA\&A}%
\def\apj{ApJ}%
\def\apjl{ApJ}%
\def\aap{A\&A}%
\def\mnras{MNRAS}%
\def\apjs{ApJS}%

\usepackage{colordvi}

\begin{document}

\title{Spectral variability of ultraluminous X-ray sources}


\author{Jari J. E. Kajava}{
  address={Department of Physical Sciences, Astronomy division, P.O.Box 3000, 90014 University of Oulu, Finland}
}

\author{Juri Poutanen}{
  address={Department of Physical Sciences, Astronomy division, P.O.Box 3000, 90014 University of Oulu, Finland}
}

\begin{abstract}
We study spectral variability of 11 ultraluminous X-ray sources (ULX) using archived {\it XMM-Newton} and {\it Chandra} observations. We use three models to describe the observed spectra; a power-law, a multi-colour disk (MCD) and a combination of these two models. We find that out of the 11 ULXs in our sample, 7 ULXs show a correlation between the luminosity and the photon index $\Gamma$ (hereafter $L-\Gamma$ correlation). Furthermore, out of the 7 ULXs that have the $L-\Gamma$ correlation, 4 ULXs also show spectral pivoting in the observed energy band. We also find that two ULXs show an $L-\Gamma$ anti-correlation. The spectra of 4 ULXs in the sample can be adequately fitted with a MCD model. We compare these sources to known black hole binaries (BHB) and find that they follow similar paths in their luminosity-temperature (hereafter $L-T$) diagrams. Finally we show that the 'soft excess' reported for many of these ULXs at $\sim 0.2$ keV seem to follow a trend $L \propto T^{-4}$ when modeled with a power-law plus a 'cool' MCD model. This is contrary to the expected $L \propto T^{4}$ relation that is expected from theory and what is seen for many accreting BHBs. 
\end{abstract}

\maketitle


\section{Introduction}

Ultraluminous X-ray sources (ULX) are point like, non-nuclear (not an active galactic nucleus, AGN) objects with X-ray luminosities exceeding the Eddington limit for stellar mass black holes (StMBH) ($L_X \gtsim 2 \cdot 10^{39}$ erg s$^{-1}$). Many ULXs have been identified as accreting black holes because of their short- and long-term variability \citep[][and references therein]{MC04}. More than 150 ULXs have been discovered in nearby galaxies \citep{CP02,LB05} and all types of galaxies have ULXs \citep{M06}. Although they were discovered with the Einstein mission almost thirty years ago \citep[see][for review]{F89}, the nature of their X-ray emission mechanisms is not yet understood. 

By assuming that ULXs are isotropic emitters, the most luminous source M82 X-1 has a luminosity of $L \sim 10^{41} \, \ergs$ \citep{PZ06}. Therefore, if ULXs do not exceed the Eddington limit $\Ledd \approx 1.3 \cdot 10^{38} (M/M_{\odot})\, \ergs$, the accreting black hole should be an intermediate-mass black hole (IMBH) with $M \gtsim 1000 M_{\odot}$. One argument used to support the idea of IMBHs powering ULXs is the presence of a $\sim 0.2$ keV 'soft excess' in the spectra of several ULXs \citep{KCPZ03,MF03}. It is not clear, however, whether this excess really is a signature of an accretion disk or, maybe, an artifact of a complicated absorber or emission from some thermal plasma surrounding the source. Other characteristics seen in ULXs in favor of IMBH interpretation (and used against the interpretation of beaming of radiation in ULXs \citep[see however][]{BKP06}) are the low-frequency quasi-periodic oscillations detected from ULXs M82, Holmberg IX and NGC 5408 \citep{SM03,DGR06,SMW07}. 

There are also several arguments against the hypothesis that ULXs are powered by accretion onto IMBHs. The main issue is the formation of such massive objects in 'normal' stellar evolution. Black holes of masses greater than $20 M_{\odot}$ are not expected to form even from the most massive stars \citep{FK01}. Furthermore, about $30 \%$ of ULXs are associated with star formation regions \citep{GGS04} and a small number of them have been associated with short lived, luminous optical counterparts \citep{M06}. These facts can be naturally explained if ULXs are powered by accretion onto StMBHs.

There are numerous models put forward on how the luminosity of a StMBH exceeds its Eddington limit. Truly super-Eddington models include the advective 'slim' disk models \citep{JAP80,A88,Kaw03} or accretion disks with strong outflows \citep{SS73,BKP06,PLF07}. Also the photon bubble models \citep{A92,B01,FB07} can produce the high observed luminosities. Other plausible scenarios involving StMBHs include beaming of radiation by a relativistic jet \citep{KFM02,FKSB06} or geometrical beaming by an outflowing wind \citep{FM01,K01}. Because of the limited observing band of the current X-ray observatories capable of detecting ULXs ($\sim 0.1 - 10$ keV), it is not surprising that many of the above models have been successfully used to produce the observed spectra of some particular ULX. This is mainly because many previous studies have been concentrated only on modelling single observations of each ULX \citep[e.g.][]{FK05,WMR06,SRW06}. 

There are some studies where the variability of ULX spectra have also been studied \citep{FeKa06,FeKa06b,BWCR08} but the samples of ULX with multiple observations has been small. The study by \citet{FeKa06} showed that the two ULXs in NGC 1313 galaxy have different types of spectral variability. Their subsequent study \citep{FK07} also showed that the 'soft excess' in NGC 1313 ULX-2 didn't follow the expected $L \propto T^{4}$ relation that is expected from theory when modeled with a MCD.

Therefore, in order to study the spectral variability of ULXs in better detail, we have selected a sample of 11 ULXs that have been observed more that 5 times with {\it XMM-Newton} and {\it Chandra}. In this paper, we present the results of three simple models that we used to describe the observed spectra; a power-law, a multi-colour disk (MCD) and a combination of these two models. Our aim is to find correlations between the model parameters and similarities between the sources in order to answer the following questions: 1) Are ULXs powered by accretion onto IMBHs, StMBHs or are ULXs a heterogeneous class of objects? 2) What is the nature of the 'soft excess'? 3) What processes produce the observed spectra?

\section{Observations and data reduction}

We searched the data archives for high S$/$N {\it XMM-Newton} and {\it Chandra} observations of ULXs that had been observed more than 5 times. We also required that the observations were not effected by photon pile-up. There were 11 sources that met this requirement; NGC 253 X-2, NGC 253 X-4, NGC 253 X-9, IC 342 X-6, NGC 1313 ULX-1, NGC 1313 ULX-2, Holmberg II ULX-1, Holmberg IX ULX-1, NGC 5204 ULX-1, NGC 5408 ULX-1 and NGC 6946 X-6. In the cases where sources have multiple names, we followed the naming convention of \citet{LB05}. 

{\it XMM-Newton} data reduction was done using the SAS version 7.1.0. In order to produce a homogeneously calibrated data set, we reprocessed all the observations using the observation data files (ODF) with the latest calibration files as of March 2008. The reduction was done according to the User's guide to the {\it XMM-Newton} Science Analysis System.\footnote{Guide is available at \\ http://xmm.esac.esa.int/external/xmm\_user\_support/documentation/sas\_usg/USG.pdf.} Specifically, we used the event selections "FLAG$==$0 \&\& PATTERN$<=$4" and "\#XMMEA\_EM \&\& PATTERN$<=$12" for the EPIC-pn and EPIC-mos instruments, respectively. {\it Chandra} observations were reduced using the CIAO 4.0.1 software. We reprocessed all the observations with calibration files from CALDB 3.4.3 following the {\it Chandra} science threads.\footnote{Threads are available at http://cxc.harvard.edu/ciao/threads/index.html.} 

We used circular source extraction regions of varying radii (between 20'' to 40'' for {\it XMM-Newton} and 4'' to 12'' for {\it Chandra}) that covered the full PSF of the sources. Close-by source free regions\footnote{In the case of {\it XMM-Newton}, the background region was chosen from the same EPIC-pn CCD chip as the source.} were used as a background. The resulting background subtracted spectral data were binned to require at least 20 counts per bin to ensure adequate statistic for the XSPEC spectral fitting. To ensure the best possible spectral quality, the time periods of high background were eliminated from the data for both {\it XMM-Newton} and {\it Chandra} observations.

\section{Spectral analysis and results}

We used three models to describe the observed spectra; a power-law, a MCD ({\sc diskbb} model in XSPEC) and a combination of these two models. The effect of the interstellar medium was taken into account with the {\sc wabs} model in XSPEC. The hydrogen column density was let as a free parameter but was required to have at least the galactic value \citep{DL90}. 

All the parameter errors in the following figures are given at the $90\%$ confidence level. The errors of the intrinsic luminosities were estimated by repeatedly setting one of the free parameters into their minimum (or maximum) values and then fitting the rest of the free parameters. We then set the absorption column to zero,  computed the intrinsic luminosity, and selected the minimum and maximum luminosities as the estimate for the error. In the case of the power-law, the luminosities were calculated for the $0.4-10$ keV band. The MCD component luminosities are bolometric.

\subsection{Power-law type ULXs and the $L-\Gamma$ correlation}

\begin{figure}[tbp]
\includegraphics[width=\linewidth]{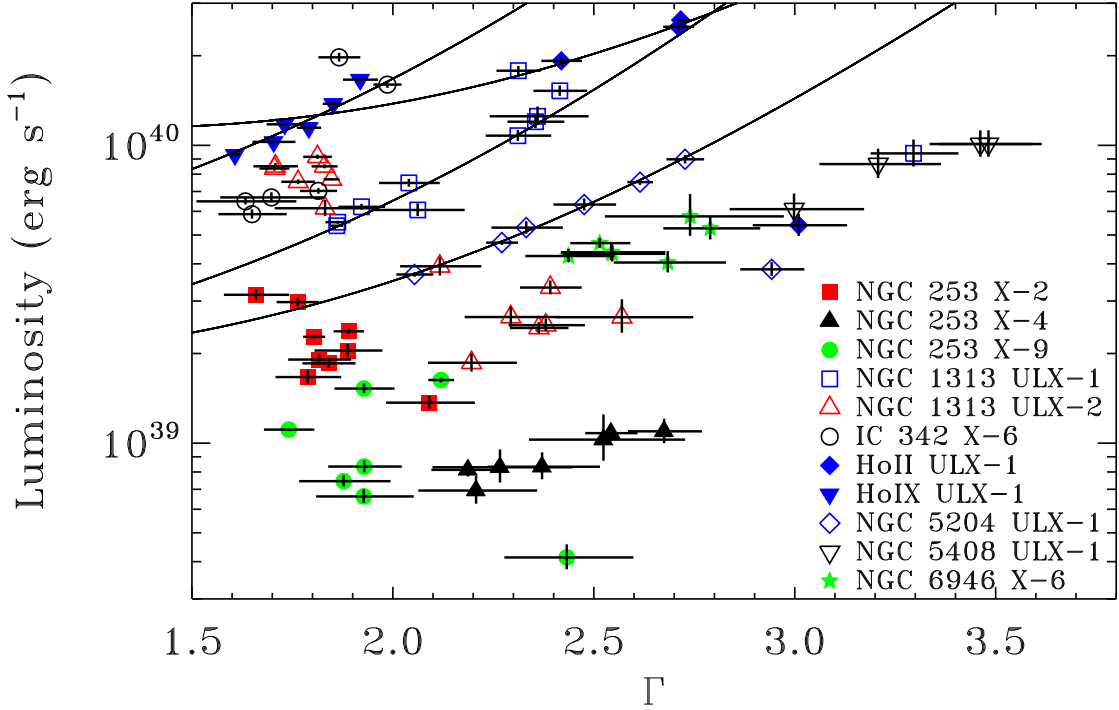}
\caption{The intrinsic $0.4-10$ keV band luminosity versus the photon index. Various observations of a given objects showing spectral pivoting in the observed band are connected by a theoretical curve (\ref{Fgamma}). Some sources show correlation, but no spectral pivoting and some other even show anti-correlation. NGC 1313 ULX-1, Holmberg II ULX-1, Holmberg IX ULX-1, NGC 5204 ULX-1 show $L-\Gamma$ correlation {\em with} spectral pivoting in the observed band. $L-\Gamma$ correlated sources {\em without} spectral pivoting are NGC 253 X-4, IC 342 X-6 and NGC 5408 ULX-1. NGC 253 X-2 and NGC 1313 ULX-2 are $L-\Gamma$ anti-correlated sources. NGC 253 X-9 shows no $L-\Gamma$ correlation and NGC 6946 X-6 is not variable enough to determine any correlations.}
\label{pofitresults}
\end{figure}

In Fig. \ref{pofitresults} we show the results of fitting with the power-law model. There is a correlation between the luminosity and the photon index $\Gamma$ (hereafter $L-\Gamma$ correlation) for most of the ULXs in our sample. This was also observed for NGC 1313 ULX-1 by \citet{FeKa06}. The ULXs with the $L-\Gamma$ correlation are NGC 253 X-4, IC 342 X-6, NGC 1313 ULX-1, Holmberg II ULX-1, Holmberg IX ULX-1, NGC 5204 ULX-1 and NGC 5408 ULX-1. In contrast, two ULXs of the sample, NGC 253 X-2 and NGC 1313 ULX-2, show an anti-correlation between the luminosity and $\Gamma$ (see Fig. \ref{pofitresults} and \citep[][fig. 2]{FeKa06}). We also find that the luminosity of NGC 253 X-9 is clearly not correlated with $\Gamma$ and that NGC 6946 X-6 is not significantly variable to determine a possible correlation.

In general, all the sources that have luminosities {\em permanently above} $\sim 3 \cdot 10^{39} \, \ergs$ have power-law type spectra. In contrast, the three sources that have luminosities {\em permanently below}  $\sim 10^{39} \, \ergs$ have non-power-law (or thermal) type spectra. An exception  to this 'rule' is NGC 1313 ULX-2 which has non-power-law type spectra at high luminosities and power-law type spectra at lower luminosities very much like the very high state (VHS) spectra of XTE J1550-564 \citep{KD04}.

Out of the 7 ULXs with the $L-\Gamma$ correlation, 4 ULXs (NGC 1313 ULX-1, Holmberg II ULX-1, Holmberg IX ULX-1 and NGC 5204 ULX-1) show also pivoting in their spectral variability with pivot energies $\epiv \sim 9, 3.5, 10, 5.5$ keV, respectively. Similar pivoting has been observed e.g. in Cyg X-1 \citep{Z02}, and several AGNs \citep[see e.g.][table 1, and references therein]{Z03}. In Fig. \ref{pofitresults} we also plot the dependence between the luminosity and the photon index using the expression from \citep[][eq. 7]{Z03}
\be \label{Fgamma}
L_{E_1 - E_2} = 4\pi D^2 F_{E_1 - E_2} = C E_p^{\Gamma} \frac{E_2^{2 - \Gamma} - E_1^{2 - \Gamma}}{2-\Gamma},
\ee
where $D$ is the distance, $E_1$ and $E_2$ are the energy band values, $F_{E_1 - E_2}$ is the flux in the selected energy band and $C$ is a constant. Eq. (\ref{Fgamma}) give a good representation of the observed $L-\Gamma$ relation for all the pivoting ULXs. However, NGC 1313 ULX-1, Holmberg II ULX-1 and NGC 5204 ULX-1 also show deviations from this when the spectrum gets very soft, i.e. when $\Gamma \gtsim 3$ keV.

\begin{figure}[tbp]
\begin{tabular}{c c}
 \includegraphics[width=0.5\linewidth]{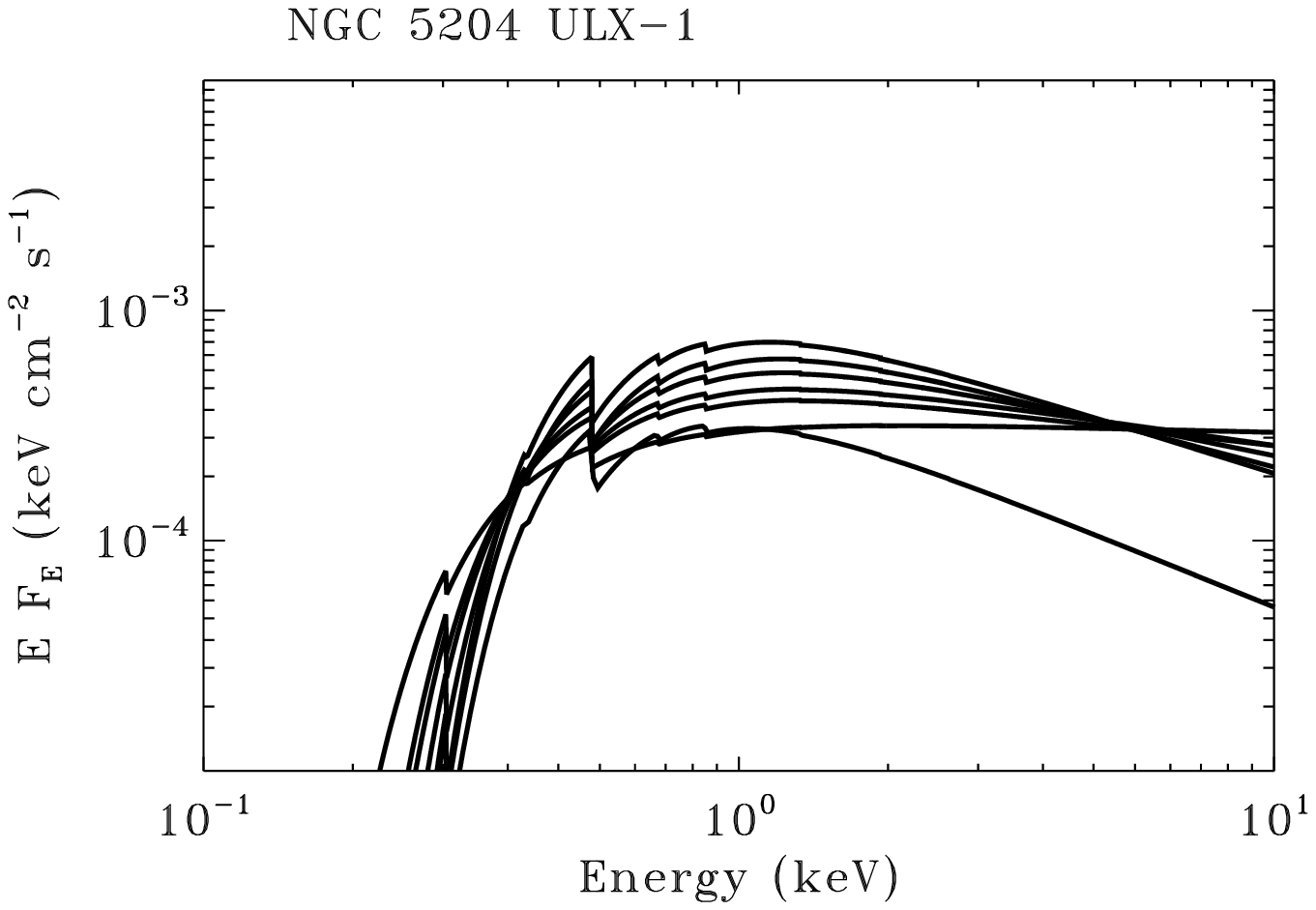}
 & \includegraphics[width=0.5\linewidth]{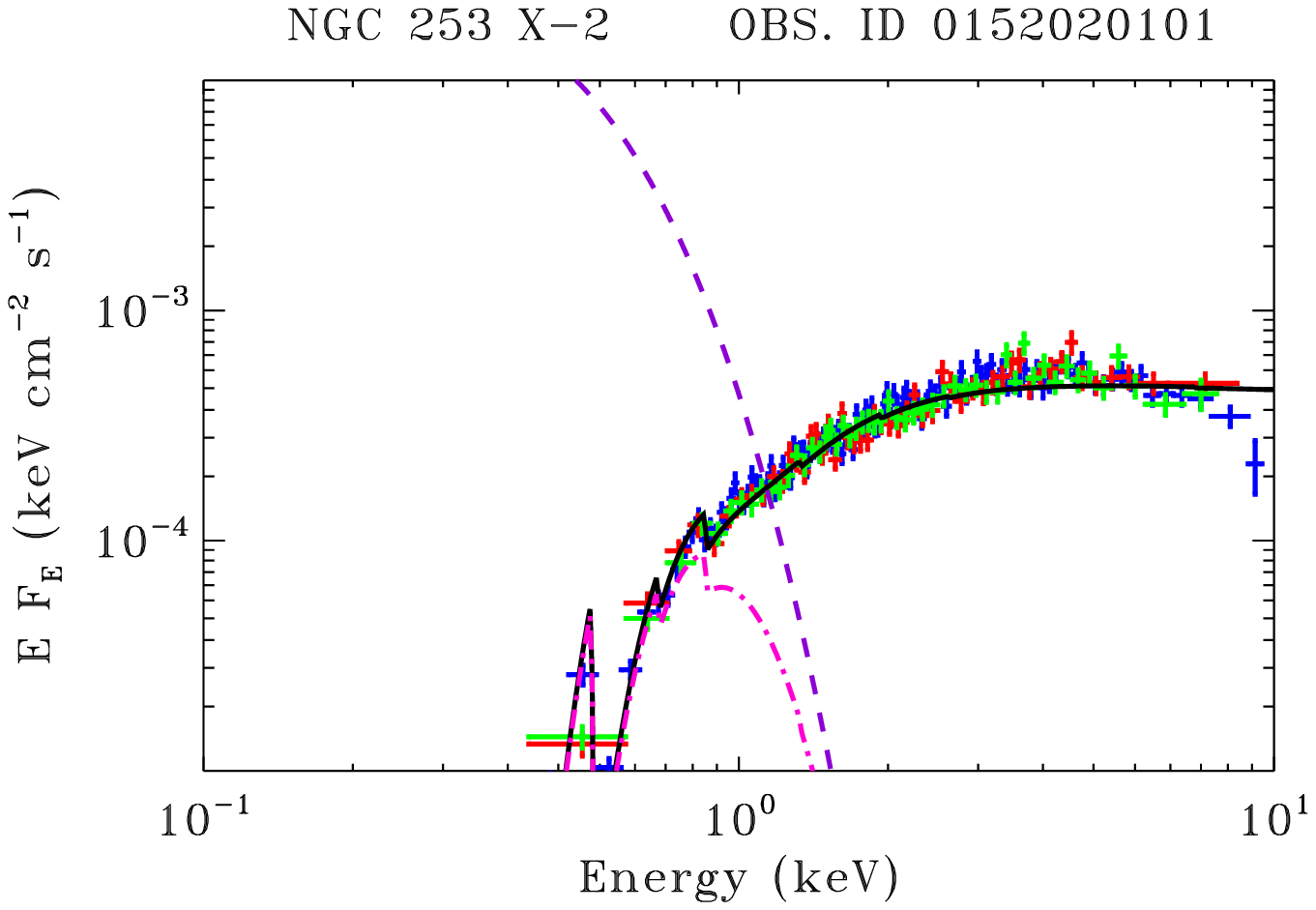} \\
\caption{Left: All observed $EF_E$ spectra of NGC 5204 ULX-1 fitted with an absorbed power-law model. The curves show the model spectra. Spectral pivoting occurs at $\sim 5.5$ keV. Right: $EF_E$ spectra of NGC 253 X-2 fitted with an absorbed power-law plus MCD model. EPIC pn, mos1 and mos2 data are shown in blue, red and green, respectively.  
The absorbed MCD component is shown with a pink dot-dashed line and the dashed purple line shows the MCD component corrected for absorption. A simple absorbed MCD fit with $\Tin \sim 1.5$ keV improves the fit from $\chired \sim 1.17$ to $\chired \sim 1.08$. Clearly for this observation the highly luminous 'cool' $\sim 0.1$ keV MCD component results from the improper modeling of the underlying power-law component, which leads to almost an order of magnitude increase in the modeled absorption column density.}
\label{spectra}
\end{tabular}
\end{figure}

The process producing the observed power-law type spectrum is most likely Compton upscattering of soft disk photons in a hot plasma surrounding the disk. The observed $L-\Gamma$ correlation and the pivoting in the sources can also be explained by variability of the luminosity of the soft disk photons. A larger disk luminosity leads to a softer and more luminous X-ray spectrum \citep{Z03}. An example of such variability is for a constant hot-plasma luminosity is given in fig. 3 of \citet{ZG01} for 3C 120. The similarity of this source to NGC 5204 ULX-1 (and the other 3 ULXs with pivoting) is remarkable (see Fig. \ref{spectra}). Unfortunately the limited observing band of {\it XMM-Newton} and {\it Chandra} do not allow us to see the high energy cutoff of the spectrum which could confirm the assumption of the constant hot-plasma luminosity. The deviations from the $L-\Gamma$ correlation by the three pivoting ULXs can be also understood by a change in the mass accretion rate \cite[see][fig 14c]{Z02}.

\subsection{Non-power-law type ULXs}

\begin{figure}[tbp]
\begin{tabular}{c c}
 \includegraphics[width=0.5\linewidth]{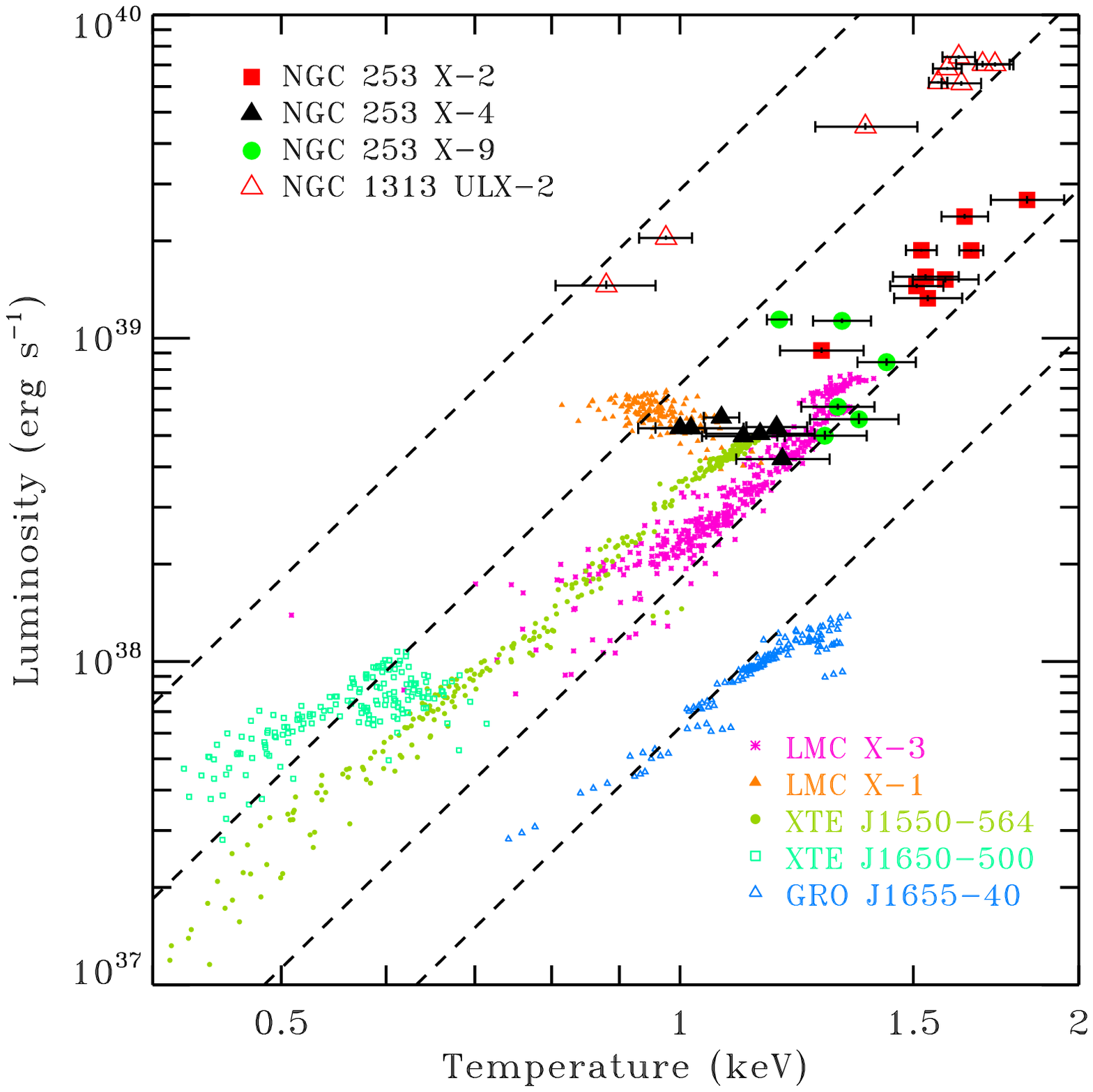}
 & \includegraphics[width=0.5\linewidth]{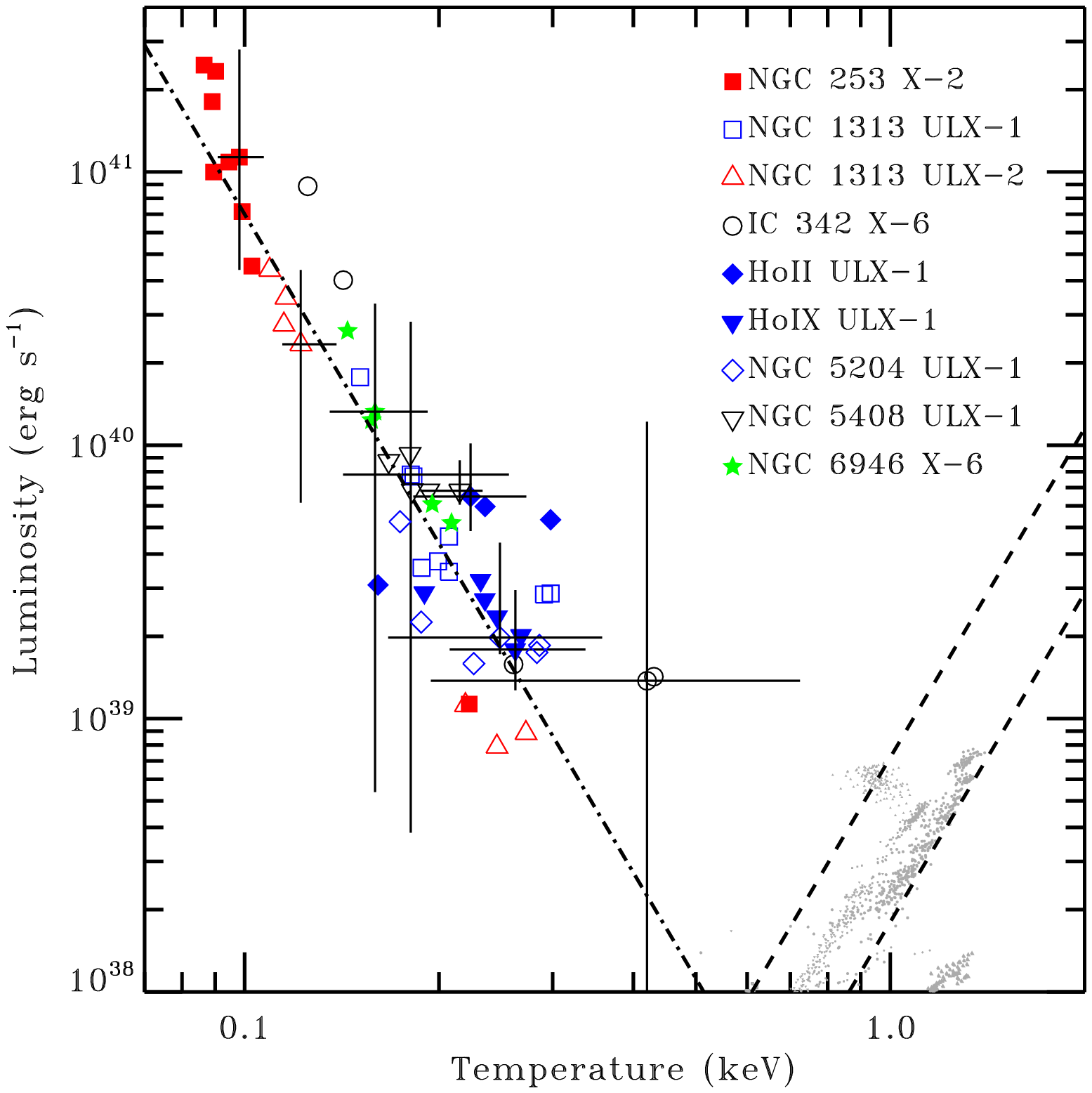} \\
\caption{Left: The luminosity-temperature diagram for the 'non-power-law type' ULXs. All the sources follow the tracks of known BHBs. Right: The luminosity-temperature diagram of the 'soft excess'. All the sources that can be modeled with the power-law plus cool MCD model follow roughly a $L \propto \Tin^{-4}$ scaling. Note however that the coolest and highest luminosity 'soft excesses' are artifacts of an improper modeling of the hard emission (see Fig. \ref{spectra}).}
\label{diskfitresults}
\end{tabular}
\end{figure}

Out of the  selected 11 ULXs, four ULXs can be adequately fitted with an absorbed MCD model. These sources include the two ULXs that show $L-\Gamma$ anti-correlation (NGC 253 X-2 and NGC 1313 X-2) and the two low luminosity ULXs (NGC 253 X-4 and NGC 253 X-9) of the sample. The resulting fits are shown in a luminosity-temperature (hereafter $L-T$) diagram in Fig. \ref{diskfitresults} together with the observations on nearby known black hole binaries adopted from \citet{GD04}. Clearly,  these four ULXs follow the behaviour seen in BHBs. The properties of the sources can be summarized as follows.  

(i) NGC 253 X-2 closely follows the $L \propto T^{4}$ relation. Its track on the diagram is a high luminosity--high temperature extension to XTE J1550-564 and LMC X-3. Its relatively high luminosity compared to BHB and modeled hot inner disk temperatures could be explained by assuming that this source is a maximally rotating Kerr black hole with $M \sim 70 M_{\odot}$ and inclination of $\sim 70$ degrees \citep{HK08}.

(ii) NGC 253 X-4 on the other hand behaves like LMC X-1 which clearly does not follow the expected $L \propto T^{4}$ relation. As discussed in \citet{PLF07}, this can be understood if the sources are very low mass black holes or even neutron stars accreting at slightly above their Eddington accretion rates. In this case the observed temperature corresponds to the emission from the spherization radius or even the distant photosphere if observed edge-on.

(iii) NGC 253 X-9 first seems to follow the $L \propto T^{4}$ track at lower luminosities, but then at $L \gtsim 10^{39} \, \ergs$ deviates towards lower temperatures similarly to LMC X-1 and NGC 253 X-4. This type of behaviour is also seen in the spectra of XTE J1550-564 when it is in the VHS instead of the high/soft state \citep{KD04}.

(iv) NGC 1313 X-2 shows a slight deviation from the $L \propto T^{4}$ track much like XTE J1650-500. As the luminosity of this source greatly exceeds the Eddington limit for a 10 solar mass black hole, the simple {\sc diskbb} model is not anymore a good approximation of the spectrum. Indeed the {\sc diskbb} model failed to properly model the high-energy curvature of the spectrum, predicting slightly too steep cutoff. Because at lower luminosities the observed spectra of this source is very much like VHS spectra of XTE J1550-564 \citep{KD04}, the higher luminosity non-power-law type spectra could be understood as a 'superluminous' state. However, a use of a more elaborate model than the {\sc diskbb} should be used to investigate this state further.

\subsection{The 'soft excess' and the nature of ULXs}

We show a luminosity-temperature relation for the 'soft excess' in the right panel of Fig. \ref{diskfitresults}. Here the model is a combination of a power-law and the {\sc diskbb} model. We find that the intrinsic luminosity of the MCD component roughly scales as $L \propto \Tin^{-4}$. However, this result should be taken with some caution. The highly luminous low temperature excesses ($L \gtsim 10^{40}\, \ergs$, $\Tin \sim 0.1$ keV) do not seem to be real features but rather artifacts of improper  modeling of the harder emission with a power-law. This results in a rise in the modeled absorption column which then causes the increase in the estimated intrinsic luminosity of the 'soft excess' (see Fig. \ref{spectra}). Never the less, the excesses below $L \ltsim 10^{40}\, \ergs$, $\Tin \gtsim 0.1$ keV seem to be real features, but the observed trend is not consistent with the interpretation of an accretion disk around an IMBH.

If the observed soft excess does not originate from a disk around an IMBH, what is then its origin? One possible answer could be the 'super-Eddington outflow' model by \citet{PLF07} which predicts that the emission from the spherization radius 
becomes cooler with increasing luminosity. However, the observed $L \propto \Tin^{-4}$ relation is too steep for the predicted relation of this model. The ULXs that have this 'soft excess' in our sample have a power-law type spectra. This spectrum most likely arises from Comptonization of the disk photons in a surrounding corona. Therefore, a natural explanation to both the observed cool accretion disks and power-law type spectra could be the 'coupled disk-corona' model \citep{DK06,SD06}. In this model the corona drains power from the underlying disk which then effectively cools and becomes less luminous. However, there is one important question that this model fails to answer. Assuming that the ULXs are at $\sim$ Eddington luminosity sources with masses $\sim 100 M_{\odot}$, why would the sources be permanently in this 'coronal dominant' state? Or in other words, why are they never in a disk dominated (high/soft) state with inner disk temperatures of $\sim 0.5$ keV?

Therefore, it is very plausible that many of the above mechanisms operate simultaneously and therefore the ULX variability results from an interplay between outflows, advection, beaming (geometrical or relativistic) which is then affected by Comptonization. One possible scenario explaining the nature of the observed spectral variability could be a 'hybrid' model combining the 'outflow' model  and the 'coupled disk-corona' model. At high inclinations the central Comptonized part of the disk would be obscured by the outflow. This would result in a non-power-law looking spectra that we see for NGC 1313 ULX-2 at high luminosities and for all the NGC 253 ULXs. On the other hand, at low inclination angles the central hot corona would be visible to the observer at all times and the outflowing wind would cause the Comptonized emission to be geometrically beamed, thus resulting in a high $\sim 10^{40} \, \ergs$ apparent luminosity. However, it could be evenly plausible that we see power-law and non-power-law type ULXs and different kind of variability simply because ULXs are a heterogeneous class of objects.

\section{Summary and conclusions}

We have studied spectral variability of 11 ULXs and  can draw several conclusions.
\begin{itemize}
\item The ULXs of this sample can be roughly separated into two types based on their spectra: power-law- and non-power-law type ULXs. All the power-law type ULXs are constantly more luminous than $\sim 3 \cdot 10^{39} \, \ergs$, i.e. they show permanently  super-Eddington luminosities for a $\sim 10 M_{\odot}$ black hole.

\item Six out of the seven power-law type ULXs show a $L-\Gamma$ correlation, i.e. the observed spectrum softens as the luminosity increases. Furthermore, out of the 7 ULXs with the $L-\Gamma$ correlation, 4 ULXs  show also pivoting in their spectral variability with $\epiv \ltsim 10$ keV. Also one non-power-law type ULX (NGC 253 X-4) show the $L-\Gamma$ correlation. Interestingly, this source is the only one of the non-power-law type ULXs that does not follow the $L \propto T^{4}$ track when modeled with a MCD.

\item The non-power-law type ULXs can be adequately modeled with a MCD. These sources also follow the behaviour of known BHBs in the $L-T$ diagrams. 

\item NGC 1313 ULX-2 is a very interesting source showing a VHS type spectra (similar to XTE J1550-564) at lower luminosities and a non-power-law (thermal) like spectra at high luminosities. It is the only source in this sample showing transitions between a spectral state that is seen to known BHB and a 'hyperluminous' spectral state.

\item The 'soft excess' seems to follow a $L \propto T^{-4}$ relation. Furthermore, the highly luminous low-temperature excesses ($L \gtsim 10^{40} \ergs$, $\Tin \sim 0.1$ keV) do not seem to be real features but rather artifacts of improper modeling of the harder emission. This $L \propto T^{-4}$ relation does not support the idea that the observed soft excess arises from the disk emission around IMBHs.
\end{itemize}

 The main question remained to be answered is the origin of these two ULX types or classes. It is very plausible that the two observed types arise simply because there are two different emission mechanisms producing high luminosities in ULXs. If the ULXs have outflows which cause beaming of radiation from the inner parts of the accretion disk, the two different kinds of variability can also result from different inclination angles between the two types.


\begin{theacknowledgments}

 JJEK acknowledges the Finnish Graduate School in Astronomy and Space Physics. JP 
 thanks the  Academy of Finland for support (grant 110792).

\end{theacknowledgments}






\begin{thebibliography}{42}
\expandafter\ifx\csname natexlab\endcsname\relax\def\natexlab#1{#1}\fi
\providecommand{\enquote}[1]{``#1''}
\expandafter\ifx\csname url\endcsname\relax
  \def\url#1{\texttt{#1}}\fi
\expandafter\ifx\csname urlprefix\endcsname\relax\def\urlprefix{URL }\fi
\providecommand{\eprint}[2][]{\url{#2}}

\bibitem[{Miller} and {Colbert}(2004)]{MC04}
M.~C. {Miller}, and E.~J.~M. {Colbert}, \emph{International Journal of Modern
  Physics D} \textbf{13}, 1--64 (2004). 

\bibitem[{Colbert} and {Ptak}(2002)]{CP02}
E.~J.~M. {Colbert}, and A.~F. {Ptak}, \emph{\apjs} \textbf{143}, 25--45 (2002).

\bibitem[{Liu} and {Bregman}(2005)]{LB05}
J.-F. {Liu}, and J.~N. {Bregman}, \emph{\apjs} \textbf{157}, 59--125 (2005).

\bibitem[{Mushotzky}(2006)]{M06}
R.~{Mushotzky}, \emph{Advances in Space Research} \textbf{38}, 2793--2800
  (2006).

\bibitem[{Fabbiano}(1989)]{F89}
G.~{Fabbiano}, \emph{\araa} \textbf{27}, 87--138 (1989).

\bibitem[{Patruno} et~al.(2006)]{PZ06}
A.~{Patruno}, S.~{Portegies Zwart}, J.~{Dewi}, and C.~{Hopman}, \emph{\mnras}
  \textbf{370}, L6--L9 (2006). 

\bibitem[{Kaaret} et~al.(2003)]{KCPZ03}
P.~{Kaaret}, S.~{Corbel}, A.~H. {Prestwich}, and A.~{Zezas}, \emph{Science}
  \textbf{299}, 365--368 (2003). 

\bibitem[{Miller} et~al.(2003)]{MF03}
J.~M. {Miller}, G.~{Fabbiano}, M.~C. {Miller}, and A.~C. {Fabian}, \emph{\apjl}
  \textbf{585}, L37--L40 (2003).

\bibitem[{Begelman} et~al.(2006)]{BKP06}
M.~C. {Begelman}, A.~R. {King}, and J.~E. {Pringle}, \emph{\mnras}
  \textbf{370}, 399--404 (2006). 

\bibitem[{Strohmayer} and {Mushotzky}(2003)]{SM03}
T.~E. {Strohmayer}, and R.~F. {Mushotzky}, \emph{\apjl} \textbf{586}, L61--L64
  (2003). 

\bibitem[{Dewangan} et~al.(2006)]{DGR06}
G.~C. {Dewangan}, R.~E. {Griffiths}, and A.~R. {Rao}, \emph{\apjl}
  \textbf{641}, L125--L128 (2006). 

\bibitem[{Strohmayer} et~al.(2007)]{SMW07}
T.~E. {Strohmayer}, R.~F. {Mushotzky}, L.~{Winter}, R.~{Soria}, P.~{Uttley},
  and M.~{Cropper}, \emph{\apj} \textbf{660}, 580--586 (2007).

\bibitem[{Fryer} and {Kalogera}(2001)]{FK01}
C.~L. {Fryer}, and V.~{Kalogera}, \emph{\apj} \textbf{554}, 548--560 (2001).

\bibitem[{Gilfanov} et~al.(2004)]{GGS04}
M.~{Gilfanov}, H.-J. {Grimm}, and R.~{Sunyaev}, \emph{\mnras} \textbf{347},
  L57--L60 (2004). 

\bibitem[{Jaroszy{\'n}ski} et~al.(1980)]{JAP80}
M.~{Jaroszy{\'n}ski}, M.~A. {Abramowicz}, and B.~{Paczy{\'n}ski}, \emph{Acta
  Astronomica} \textbf{30}, 1--34 (1980).

\bibitem[{Abramowicz} et~al.(1988)]{A88}
M.~A. {Abramowicz}, B.~{Czerny}, J.~P. {Lasota}, and E.~{Szuszkiewicz},
  \emph{\apj} \textbf{332}, 646--658 (1988).

\bibitem[{Kawaguchi}(2003)]{Kaw03}
T.~{Kawaguchi}, \emph{\apj} \textbf{593}, 69--84 (2003). 

\bibitem[{Shakura} and {Sunyaev}(1973)]{SS73}
N.~I. {Shakura}, and R.~A. {Sunyaev}, \emph{\aap} \textbf{24}, 337--355 (1973).

\bibitem[{Poutanen} et~al.(2007)]{PLF07}
J.~{Poutanen}, G.~{Lipunova}, S.~{Fabrika}, A.~G. {Butkevich}, and
  P.~{Abolmasov}, \emph{\mnras} \textbf{377}, 1187--1194 (2007).

\bibitem[{Arons}(1992)]{A92}
J.~{Arons}, \emph{\apj} \textbf{388}, 561--578 (1992).

\bibitem[{Begelman}(2001)]{B01}
M.~C. {Begelman}, \emph{\apj} \textbf{551}, 897--906 (2001).

\bibitem[{Finke} and {B{\"o}ttcher}(2007)]{FB07}
J.~D. {Finke}, and M.~{B{\"o}ttcher}, \emph{\apj} \textbf{667}, 395--403
  (2007). 

\bibitem[{K{\"o}rding} et~al.(2002)]{KFM02}
E.~{K{\"o}rding}, H.~{Falcke}, and S.~{Markoff}, \emph{\aap} \textbf{382},
  L13--L16 (2002).

\bibitem[{Freeland} et~al.(2006)]{FKSB06}
M.~{Freeland}, Z.~{Kuncic}, R.~{Soria}, and G.~V. {Bicknell}, \emph{\mnras}
  \textbf{372}, 630--638 (2006). 

\bibitem[{Fabrika} and {Mescheryakov}(2001)]{FM01}
S.~{Fabrika}, and A.~{Mescheryakov}, 
in \emph{IAU Symp. 205,  Galaxies and their Constituents at the Highest Angular Resolution}, eds.
  R.~T. {Schilizzi}, S.~N. {Vogel}, F.~{Paresce}, and M.~S. {Elvis}, Astron.
  Soc. Pac., San Francisco, 2001, pp. 268--269. 

\bibitem[{King} et~al.(2001)]{K01}
A.~R. {King}, M.~B. {Davies}, M.~J. {Ward}, G.~{Fabbiano}, and M.~{Elvis},
  \emph{\apjl} \textbf{552}, L109--L112 (2001).

\bibitem[{Feng} and {Kaaret}(2005)]{FK05}
H.~{Feng}, and P.~{Kaaret}, \emph{\apj} \textbf{633}, 1052--1063 (2005).

\bibitem[{Winter} et~al.(2006)]{WMR06}
L.~M. {Winter}, R.~F. {Mushotzky}, and C.~S. {Reynolds}, \emph{\apj}
  \textbf{649}, 730--752 (2006). 

\bibitem[{Stobbart} et~al.(2006)]{SRW06}
A.-M. {Stobbart}, T.~P. {Roberts}, and J.~{Wilms}, \emph{\mnras} \textbf{368},
  397--413 (2006).

\bibitem[{Feng} and {Kaaret}(2006{\natexlab{a}})]{FeKa06}
H.~{Feng}, and P.~{Kaaret}, \emph{\apjl} \textbf{650}, L75--L78
  (2006{\natexlab{a}}). 

\bibitem[{Feng} and {Kaaret}(2006{\natexlab{b}})]{FeKa06b}
H.~{Feng}, and P.~{Kaaret}, \emph{\apj} \textbf{653}, 536--544
  (2006{\natexlab{b}}). 

\bibitem[{Berghea} et~al.(2008)]{BWCR08}
C.~T. {Berghea}, K.~A. {Weaver}, E.~J.~M. {Colbert}, and T.~P. {Roberts},
\eprint{0807.1547}.

\bibitem[{Feng} and {Kaaret}(2007)]{FK07}
H.~{Feng}, and P.~{Kaaret}, \emph{\apjl} \textbf{660}, L113--L116 (2007).

\bibitem[{Dickey} and {Lockman}(1990)]{DL90}
J.~M. {Dickey}, and F.~J. {Lockman}, \emph{\araa} \textbf{28}, 215--261 (1990).

\bibitem[{Kubota} and {Done}(2004)]{KD04}
A.~{Kubota}, and C.~{Done}, \emph{\mnras} \textbf{353}, 980--990 (2004).

\bibitem[{Zdziarski} et~al.(2002)]{Z02}
A.~A. {Zdziarski}, J.~{Poutanen}, W.~S. {Paciesas}, and L.~{Wen}, \emph{\apj}
  \textbf{578}, 357--373 (2002).

\bibitem[{Zdziarski} et~al.(2003)]{Z03}
A.~A. {Zdziarski}, P.~{Lubi{\'n}ski}, M.~{Gilfanov}, and M.~{Revnivtsev},
  \emph{\mnras} \textbf{342}, 355--372 (2003). 

\bibitem[{Zdziarski} and {Grandi}(2001)]{ZG01}
A.~A. {Zdziarski}, and P.~{Grandi}, \emph{\apj} \textbf{551}, 186--196 (2001).

\bibitem[{Gierli{\'n}ski} and {Done}(2004)]{GD04}
M.~{Gierli{\'n}ski}, and C.~{Done}, \emph{\mnras} \textbf{347}, 885--894
  (2004).

\bibitem[{Hui} and {Krolik}(2008)]{HK08}
Y.~{Hui}, and J.~H. {Krolik}, \emph{\apj} \textbf{679}, 1405--1412 (2008).

\bibitem[{Done} and {Kubota}(2006)]{DK06}
C.~{Done}, and A.~{Kubota}, \emph{\mnras} \textbf{371}, 1216--1230 (2006).

\bibitem[{Socrates} and {Davis}(2006)]{SD06}
A.~{Socrates}, and S.~W. {Davis}, \emph{\apj} \textbf{651}, 1049--1058 (2006)

\end{thebibliography}


\end{document}